# A Permutation Test for Assessing the Presence of Individual Differences in Treatment Effects

Chi Chang – Michigan State University

Thomas Jaki – Lancaster University

Muhammad Saad Sadiq – University of Miami

Alena A. Kuhlemeier – University of New Mexico

Daniel Feaster – University of Miami

Nathan Cole – University of New Mexico

Andrea Lamont – University of South Carolina

Daniel Oberski – Utrecht University

Yasin Desai – Lancaster University

The Pooled Resource Open-Access ALS Clinical Trials Consortium*

M. Lee Van Horn – University of New Mexico

Running Head: A Permutation Test for Assessing Heterogeneity in Treatment Effects

Corresponding Author:

M. Lee Van Horn, PhD. Tech 274, 1 University of New Mexico, Albuquerque, NM 87131

mlvh@unm.edu; (505) 277-4535

This paper was supported by grant # MR/L010658/1 awarded to Thomas Jaki by the United Kingdom Medical Research Council. For further information or comments please contact the senior author, M. Lee Van Horn at mlvh@unm.edu

Abstract

One size fits all approaches to medicine have become a thing of the past as the understanding of individual differences grows. The paper introduces a test for the presence of heterogeneity in treatment effects in a clinical trial. .Heterogeneity is assessed on the basis of the predicted individual treatment effects (PITE) framework and a permutation test is utilized to establish if significant heterogeneity is present. We first use the novel test to show that heterogeneity in the effects of interventions exists in the Amyotrophic Lateral Sclerosis Clinical Trials. We then show, using two different predictive models (linear regression model and Random Forests) that the test has adequate type I error control. Next, we use the ALS data as the basis for simulations to demonstrate the ability of the permutation test to find heterogeneity in treatment effects as a function of both effect size and sample size. We find that the proposed test has good power to detected heterogeneity in treatment effects when the heterogeneity was due primarily to a single predictor, or when it was spread across the predictors. The predictive model, on the other hand is of secondary importance to detect heterogeneity. The non-parametric property of the permutation test can be applied with any predictive method and requires no additional assumptions to obtain PITEs.

1. Introduction

The key premise of precision medicine is the identification and targeting of individuals most likely to benefit from a given intervention,[1] with the goal of improving health care outcomes and decreasing costs.[1,2] Much recent research has focused on statistical approaches for identifying a small number of subgroups of individuals who differ in their response to interventions,[3–13] while a smaller body of research has focused on predicting intervention responses at an individual level.[4,14–19] For situations in which treatment response is related to a set of covariates which is not small number of clearly defined subgroups, individual-level predictions are particularly appropriate. Even if most covariates were categorical, with high dimensional data and finite samples individual level predictions may contain more information about heterogeneity in treatment effects than is contained in subgroups. This study focuses on the use of predicted individual treatment effects (PITE),[20,21] a framework based on potential outcomes[22,23] that results in predictions of responses interventions tailored to each individual patient.

The PITE approach utilizes data from a randomized clinical trial with a potentially very large number of baseline covariates to generate predictions from a model or algorithm, which are then used in estimating PITEs. The same model or algorithm can then be used to generate treatment effect estimates for new subjects not used in training. Given that predictive algorithms have been trained, the next question becomes whether the data reveal more variability in individual predictions than would be expected due to chance. In other words, 'Do individuals differ in the effects of the intervention?' is a question that should be answered before PITE estimates from a given trial are used because otherwise the PITE predictions provide no information beyond the average treatment effect. This paper proposes a permutation test to answer this question using predictions from the PITE framework. The null hypothesis of the permutation test is that the PITE predictions explain no more variance than using average treatment effect (ATE). An advantage of the proposed method is that it can be generally applied to any method for estimating predictions for the treated group and the control group. While methods exist for estimating the significance of heterogeneity in treatment effects using kernel regression and instrumental variable regression,[24,25] the proposed permutation test provides flexibility in choosing the estimator and can use machine learning for the two potential outcomes while retaining frequentist properties. The next section describes the PITE approach in general terms before providing details of our proposed permutation test. In Section 3, we use the PITE framework and the proposed test to evaluate heterogeneity in the effects of interventions for ALS; Section 4 uses this test on simulated data to show the type I error rates of the PITE permutation test using two different predictive models with and without main effects of covariates. In Section 5, we use it as the basis for simulations that demonstrate the ability of the permutation test to find heterogeneity in treatment effects as a function of both effect size and sample size. Section 6 concludes with a discussion of results.

2. Permutation test for PITE

In a clinical trial, we observe the outcome for a given patient under either the experimental or the control condition. This has been highlighted in the causal inference literature[23,26,27] and leads to challenges when one aims to estimate patient-level treatment effects. The ATE is usually defined as

$$\text{ATE} = E(Y_1) - E(Y_0) \qquad (1)$$

where $E(Y_1)$ is the expected response under experimental treatment and $E(Y_0)$ the expected response under control, possibly also adjusting for covariates. Heterogeneity in treatment effects implies that there is individual variability in the ATE such that some individuals are expected to do better than average, and some are expected to do worse. It should be noted that when the ATE equals zero it is still possible that there are some individuals who would be expected to do better given the treatment than control and others who would be expected to do better under control. Therefore, in this paper, we exclude the expected value of the PITE from the test, as this value is equal to the ATE and not evidence of individual differences. Lamont et al.[20] defines PITE as the difference between the potential (or counterfactual predicted) outcome under treatment (t) and control (c) for each patient *i* given their observed covariates ***X***.

$$\text{PITE}_i = Y_i^t - Y_i^c \tag{2}$$

where $Y_i^t$ indicates each patient's potential outcome if they all get the treatment, and $Y_i^c$ is patient's potential outcome if they all in the control condition. Then the difference between ATE in Equation (1) and PITE framework in Equation (2) is that PITE focuses on individual's potential outcome therefore, it can be used with any predictive model that allows outcome prediction on a patient level (e.g., random forests,[28] Bayesian additive regression trees,[29] neural networks[30]). In addition, PITE can be used for predictions of treatment effects given information on covariates for patients who are not originally part of the clinical trial.

While the PITEs from Equation (2) include both the ATE and predicted individual differences in the treatment effects, the presence of individual differences has major implications for how a treatment would be implemented: if there are individual differences in the treatment effect it suggests that it may be worthwhile to collect and use individual level data to help guide treatment decisions. Therefore, we propose a permutation test to evaluate whether the individual differences observed in PITEs account for significantly more variability than the ATE alone. This paper demonstrates the use of a permutation test with two different predictive approaches, Random Forests[28] and linear regression. Since $Y_i^t$ and $Y_i^c$ in Equation (2) is not observable for any patient under both counterfactual conditions, in this framework, we estimate a predicted $Y_i^*$ as a function of observed baseline covariates: the predicted $Y_i^t$ is estimated by $f(\widehat{Y_i}|X_i, T = 1)$ and $Y_i^c$ by $f(\widehat{Y_i}|X_i, T = 0)$, where $f(.)$ indicates any predictive function.

Let y$_{jc}$ denote the observed outcome for patient *j* in the control group while y$_{kt}$ is the observed outcome for patient *k* in the experimental group. Using linear regression as an example, the outcomes for patients in the control group are regressed on their individual characteristics, **X**$_c$, via

$$y_{ic} = \mathbf{X}_c \beta_c + \varepsilon_c \tag{3}$$

which can then be used to obtain individual-level predictions of potential outcomes under control $Y_i^c$. **X**$_c$ is the design matrix, which captures all baseline covariates of the individuals in the control group. The error terms, denoted $\varepsilon_c$, are assumed to be independent and normally distributed.

Potential outcomes under the experimental condition can be estimated in the same way. Individual-level PITE estimates, both for patients in the original trial and those who did not take part in it, can then be obtained using Equation 2. It should also be noted that the algorithm or model used will determine both the assumptions made and the efficiency of the predictions (e.g., linear models assume linearity in the parameters and that all multiway interactions are included in the covariates and tends to have increased efficiency when those assumptions are met). This paper proceeds to use both Random Forests and the linear model to obtain predictions. We expect that the best model will be situation dependent.

The approach described above will yield a prediction of an individual patient's treatment effect. However, it does not provide a direct test of whether the heterogeneity across individuals observed in the PITEs is greater than chance. We therefore propose that a permutation test[23,31–34] be used to assess whether the observed differences in individual patients' PITEs are more variable than what would be expected through random chance alone. Our test focuses on the standard deviation (SD) of the PITEs, because this estimate quantifies individual differences in the predicted treatment effects. More specifically, we test the hypothesis:

$$H_0: \sigma_{PITE} = \sigma_{chance} \quad \text{versus} \quad H_A: \sigma_{PITE} > \sigma_{chance}$$

where $\sigma_{PITE}$ is the SD of the PITEs and $\sigma_{chance}$ is the SD of the distribution one would get under zero treatment effect heterogeneity. In other words, it is also the variation that would be observed if the observed individual differences in the treatment effect are only due to random chance. To test this hypothesis, the permutation test first approximates the sampling distribution of PITE's SD under the null hypothesis, i.e. when the set of covariates in the PITE prediction models have no different effect on the outcome dependent on treatment (they may be prognostic, but not predictive), and subsequently compares the observed SD from the data against the resulting distribution.

The following algorithm describes the procedure in more detail.

1) Estimate PITE models and compute $\widehat{PITE}_i$, for all $n$ individuals in the dataset using a prediction method
2) Estimate the standard deviation of the estimated PITEs as

$$\hat{\sigma}_{PITE} = \frac{1}{n-1} \sum_{i=1}^{n} (\widehat{PITE}_i - \overline{\widehat{PITE}_i})^2$$

where $\overline{\widehat{PITE}_i} = \frac{1}{n} \sum_{i=1}^{n} \widehat{PITE}_i$.

3) Randomly permute the treatment assignment of all patients in the study.
4) Estimate the PITE model and compute $\widehat{PITE}_i^p$, using the permuted data and the same prediction method as used in step 1.
5) Estimate the standard deviation, $\hat{\sigma}_{PITE}^P$ of $\widehat{PITE}_i^p$, in the same manner as in step 2.
6) Repeat steps 3 through 5, P times.
7) Obtain the p-value, $p^P$, associated with the above hypothesis as $p^P = \frac{\sum_{p=1}^{P} I(\hat{\sigma}_{PITE}^P > \hat{\sigma}_{PITE})}{P}$, with I(.) being an indicator function equal to 1 if the condition in the parenthesis is satisfied, and 0 otherwise.
8) Reject the above hypothesis at level α if $p^P < \alpha$.

The above test examines the presence of individual patient heterogeneity on the basis of the original study data only. To confirm that heterogeneity also exists in the whole population for which PITE is used to make treatment decisions (as distinct from the study population), one can use a similar procedure as the one detailed above, except that $\hat{\sigma}_{PITE}$ is estimated based on predictions from patients who were not enrolled in the original study.

For each of 1,000 replications in this study, 1,000 permutations were used in our subsequent evaluations. Following binomial arguments, this yields a .007% uncertainty in estimated p-values of which the true value should be 5%.

3. Demonstration of the permutation test: An intervention for individuals with ALS

Amyotrophic Lateral Sclerosis (ALS, also known as Motor Neuron Disease) is a neurodegenerative disorder that affects motor neurons in the brain and spinal cord. We first estimated the PITE of patients in the Pooled Resource Open-access ALS Clinical Trials database (Pro-ACT;

http://nctu.partners.org/ProACT). In 2011, Prize4Life, in collaboration with the Northeast ALS Consortium, and with funding from the ALS Therapy Alliance, formed the Pooled Resource Open-Access ALS Clinical Trials (PRO-ACT) Consortium. The data available in the PRO-ACT Database has been volunteered by PRO-ACT Consortium members. We then used the permutation test to examine whether there was heterogeneity in individual treatment effects. Pro-ACT includes information from more than 8,500 patients with ALS, each of them participated in a clinical trial and received either a placebo or an experimental treatment. Following Küffner et al,[35] we used the slope of the ALSFRS score from a repeated measures model for each patient as the primary outcome, and the 2,910 patients (1,766 in experimental treatments and 1,144 in control ones) who had complete data for 17 covariates, treatment condition, and the outcome.

To avoid overfitting the data we advocate either choosing both the predictive method to be used and the covariates for the PITEs a priori, or adjusting for the variable selection process.[36] Here, we demonstrate the permutation test on the basis of a linear model that included 7 (out of 17) covariates that were found to have significant interaction treatment. Note that in its simplest form, with a linear model, PITE captures baseline by treatment interactions, thus we used this as the criteria for variable selection. PITEs, however, are much more general than these interactions as they capture the joint effect of many predictors and, depending on the predictive method used, implicitly capture non-linear and higher order interactions. The seven covariates used for obtaining PITE estimates were: respiratory rate, systolic blood pressure, age, gender, onset location in the Limb, use of Riluzole, and Delta Flag (coded 1 if the time between patient previously used Riluzole for than a year).

*Results*

Their descriptive statistics are shown in Appendix B and C. The linear regression model's estimates for both treatment and control conditions, which are included in Appendix A, differ across conditions. These differences in estimates are what contribute to predicted individual differences in treatment effects. Figure 1 includes the permutation distribution of SDs of PITEs based on the procedure outlined above together with the observed SD from the ALS dataset, which at .127 is on the upper tail of this distribution. The p-value for the permutation test was .005, providing strong evidence for individual-level treatment heterogeneity.

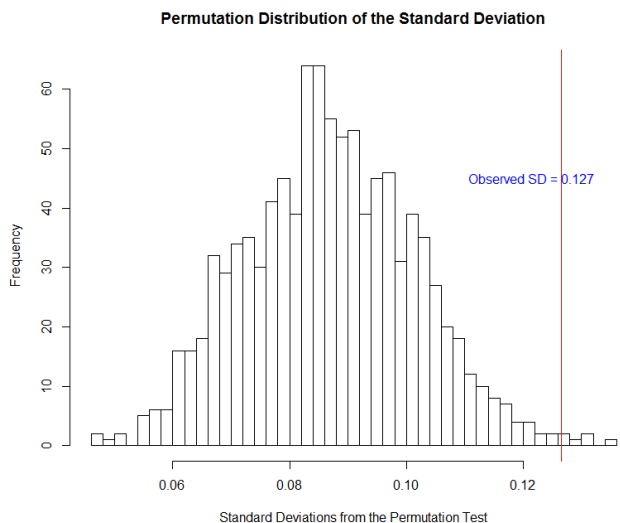

Figure 2. Permutation distribution of PITEs' SDs and the observed PITE's SD in the ALS study.

4. Type I error rates for the permutation test

The promise of the permutation test is that one can use any conventional or machine learning function and still get correct frequency properties. In this section, we investigate if this promise is fulfilled. We begin our evaluation of the proposed test by examining the type I error rate of the proposed permutation test under the null hypothesis, i.e., that the treatment effect is the same for all individuals. Simulations were conducted with the true PITE for each individual being equal to the average treatment effect, meaning that covariates had no impact on the PITE, and that $\hat{\sigma}_{PITE}$ is what would be observed if the true value of $\sigma_{PITE}$ is chance. When the type I error is .05, the p-value obtained from step 7, above, should be uniformly distributed between 0 and 1 so that the p-value is below .05 in only 5% of the simulations.

In this phase of our investigation, PITE was estimated with sample sizes of 100, 250, 500, 1,000, and 5,000 using both linear regression (via the *lm* function in R) and Random Forests (via the *randomForestSRC* package in R,[37] tuned to have a node depth of 10). To make the simulation more realistic, we included five prognostic covariates that had the same effects across the treatment and control conditions. These included three normally distributed covariates with means of 0 and variances of 1, and two binary variables, each with a 0.5 probability of endorsing 1 or 0. The covariate effects of 0.406, -0.239, 0.703, -0.090, and -0.299 respectively were identical across the treatment and the control groups – implying that they did not predict differential responses to the treatment - and were included in all analyses. To show type I error rates when many additional variables were included in the predictive model, we ran analyses with varying numbers of continuous variables with standard normal distribution and binary variables with binomial distributions with a probability of success of .5. They were generated unrelated to the outcome, hereafter called 'nuisance variables.' Because sample size limits the number of nuisance variables that can be included in a linear regression model, we examined an increased number of nuisance variables with larger samples. Analyses also varied the true treatment effect, to show that the PITE as described above does not capture main effects of treatment. For the Random Forest model, we ran with 500 trees and used 10 random split points to split a node.

*Results*

The results of such analyses (see Table 1) established that across all conditions, the permutation test rejected the null hypothesis 4.7% to 6.3 % of the time with the linear regression model, and 2.9% to 6.3% of the time with Random Forests. The estimated type I errors appear to be mostly within simulation error (±0.007) with no discernable pattern to be detectable in relation to the number of nuisance variable, main effect or total sample size.

5. Power of the permutation test

Next, we used our ALS results as the basis for simulations examining some of the factors that influence the permutation test's statistical power. In the data generation model for the power simulations, the parameters from the predictive linear model using the ALS example (shown in Appendix A) were used as the starting point. To mimic the real-world scenario, we included only the seven covariates that were found to have significant impacts on the PITE in the simulation. The first three of these, respiratory rate, systolic blood pressure, and age, were generated as normally distributed random variables with means and SDs equal to the corresponding values estimated from the ALS dataset (details of which are provided in Appendix B). Similarly, the remaining four binary random variables, gender, onset location in the limb, use of Riluzole, and Delta Flag, were generated as binomial with the same probabilities as observed (in Appendix C). The covariance matrix was generated by mimicking that of the ALS data. The outcome was generated with the effect size scaled to mimic the ALS example with sample sizes of 1,000 (to examine if the effects could have been found with a smaller sample) and 3,000 (as in the ALS example), with equal size for the treatment and placebo groups. To assess the impact of adding further covariates when fitting PITE, we evaluated statistical power with 0, 20, 50 or 100 nuisance variables, all of which were generated either from standard normal distributions or from binomial distributions with a probability of success of .5. The nuisance variables were included when estimating PITEs despite not being related to the outcome.

A challenge in estimating power was that measures of effect size for PITE have not been previously defined. In this study, we used the average PITE estimate divided by the pooled SD of the outcome as follows:

$$PITE\ Effect\ Size = \frac{\frac{\Sigma |PITE_i|}{N}}{\sqrt{\frac{(N_T - 1) \times (\sigma_T^2) + (N_C - 1) \times (\sigma_C^2)}{N_T + N_C - 1}}}$$

This resulted in an estimated effect size of 0.19 for the PITEs in the ALS example, meaning that the average person was 0.19 SD from the average effect size. When estimating power, data were generated with effect sizes of either 0.19 or 0.38, with the latter included to examine the method's ability to identify a larger effect.

We also examined the permutation test's power to detect heterogeneity that is mostly due to a single variable as well as when heterogeneity was spread across multiple variables which each contribute a small amount. Power simulations were run for six conditions, which differed from on another solely in terms of the relative contributes of the seven predictors. Specifically, these six conditions were: 1) the total heterogeneous effect is evenly spread across all seven covariates ("Spread"); 2) 90% of the total heterogeneous effect is due to the first continuous variable, and 10% to the other six covariates ("90/10 Cont."); 3) as 90/10 Cont., but with 75%/25% split between the first continuous variable and the other six covariates ("75/25 Cont."); 4) as above, but with a 50%/50% split ("50/50 Cont.");; 5) as above, but with a 25%/75% split ("25/75 Cont."); 6) 90% of the total heterogeneous effect is due to the first binary variable, and 10% to the other six covariates ("90/10 Bin."). Thus, the power of PITE prediction was examined in a total of 96 conditions, i.e., in 2 (sample sizes) × 4 (numbers of nuisance variables) × 2 (effect sizes) × 6 (heterogeneity effect distributions).

Once data was generated, both the linear model and Random Forests were run for each dataset and under each condition using the procedures described above. For the Random Forest model, the depth was restricted to 10. The percentage of times that the permutation test was significant for each condition recorded as the power estimate.

*Results*

Power for each of the 96 conditions was estimated as the proportion of 1000 simulations for which the permutation test was significant. The results obtained with an effect size of 0.19 are presented in Table 2, and those obtained with an effect size of 0.38 in Table 3. As expected, power increased both when sample size increased and when effect size increased. Our results also indicated that increasing the number of nuisance variables decreased power substantially, highlighting the importance of selecting meaningful covariates. With the ALS observed effect size and sample size, the predictive (post-hoc) power obtained from the linear regression model was adequate when there were 50 nuisance variables, but not when there were 100. When using Random Forests for predictions, on the other hand, power was adequate with 20 nuisance variables at the same sample size. However, with a sample size of 1,000, the linear regression model's power was poor even with 20 nuisance variables and would be inadequate with Random Forests using the same tuning parameters.

With an effect size twice as great as observed, the power of the permutation test using linear regression model was low only when the sample size was 1,000 and there were 50 or 100 nuisance variables, but Random Forests' power was marginal even with sample size of 3,000 if there were 100 nuisance variables. That being said, we expected that the power estimates for Random Forests would generally be lower than those for the linear regression model, given that the data were simulated using the latter and that no higher order interactions or non-linear effects were included.

The other factor that we varied across the simulations was how the effect of the covariates on heterogeneity in treatment effects was spread out. The reason for this is to show the core advantage of PITE, which is its ability to detect many small effects that add up to something meaningful rather than just one large effect. When looking across the six different spreads of the PITE effect, it was striking that when there was adequate power for one of them, there was usually adequate power for all. The only two exceptions to this were 1) when the effect was carried primarily by one binary variable (i.e., there are only two different kinds of responses), in which case, power was higher than for the other conditions; and 2) for random forests the power is lower for the binary predictor. The key result here is that, in the ALS example, power was about the same regardless of whether the heterogeneity in treatment effects is attributed primarily to one of the seven important variables or when it is spread out across all seven.

One apparently inconsistent finding in these results was that in some condition, power was less than 5%, the type I error rate. The reason for this is that any random variable will cause variability in the PITE to a certain extent. In some cases, with many nuisance variables, the effects of the predictors we were simulating were smaller than effects due to chance, resulting in a lower probability of finding heterogeneity in the effects of the predictor than would have been expected due to chance. Importantly, this implies that adding more predictors will increase the noise in PITE estimates and result in larger heterogeneity being estimated.

6. Discussion

If PITEs are to be useful for quantifying individual differences in the effects of an intervention, it is necessary to have a test that can show that differences observed are not simply due to chance. The proposed permutation test is therefore a crucial step toward rendering this approach practical. Under the 96 conditions we examined, the test was shown to have appropriate nominal type I error rates, practical utility in an applied example, and adequate power in that example given a moderately sized sample and 20 to 50 covariates. We acknowledge that the effect size observed in our applied example was fairly small (the average individual was 0.19 SD from the average treatment effect). However, if the sample size had been doubled, then the permutation test would have had adequate power even with 100 nuisance covariates. The permutation test also demonstrated an ability to detect heterogeneity in treatment effects when that is due primarily to a single predictor, or when it was spread across the seven predictors that had an impact in the ALS example. It also worked reasonably well when either the linear regression model or Random Forests were used as the predictive method. This is important because a desirable property of the permutation test is that it is a non-parametric approach which can be applied with any predictive method and makes no assumptions beyond those of the method is being used to obtain PITEs.

The permutation test was also found to have unexpected benefit, in that the variance of the PITE across permuted datasets provides an estimate of variability in PITE that, due to chance, can be attributed to unique conditions of a given application: e.g., the number and distribution of the covariates used, or the model or algorithm used to obtain predictions. Thus, this test can help assess the amount of noise added to the PITEs at a given number of covariates and a given predictive method.

We noted that for Random Forests with nuisance variables the choice of tuning parameters made a meaningful difference in the results. If there is heterogeneity in treatment effects, or added nuisance variables, the Random Forest requires more tuning or would require to be corrected for bias, irrespective of sample size. For instance, allowing deep trees led to high levels of overfitting, with many nuisance variables being identified as important. In that case, we chose a maximum node depth of 10 to reduce overfitting. However, questions of how to appropriately tune random forests when estimating PITEs are beyond the scope of this study. We mention this issue only to highlight that it is non-trivial in studies with many covariates which, while chosen because the researchers expect them to be important, may in fact be unrelated to heterogeneity in treatment effect.

The limitations of this study are that we examined the proposed PITE permutation test using predictions from only the linear regression model and Random Forests (with one set of tuning parameters), and under a set of conditions that was designed to clarify our understanding of its power via an applied example. In principle, we see no reason why this test should not work well with any method chosen but cannot claim that the present paper has established this firmly. We should also note that, in the ALS example, the permutation test required a quite large sample to attain adequate power. While the observed effect size was small in this case, even with a larger effect, the test required an N of 3,000 if many covariates were included. Because the outcome of interest is individual predictions, we believe that PITEs will generally require substantial sample sizes, unless the target effects are very large. Nevertheless, as a new approach to testing the significant of heterogeneity, the test is likely to be of considerable benefit to other similar approaches in precision medicine.

Table 1. Type 1 error rates for the PITE permutation test, the linear regression model and Random Forests.

| Sample Size | Number of Nuisance Continuous Covariates | Number of Nuisance Binary Covariates | Average Treatment Effects | Upper-sided Type I Error Rate- LM | Upper-sided Type I Error Rate- RF |
|---|---|---|---|---|---|
| 100  | 0   | 0  | 0   | 0.049 | 0.046 |
| 100  | 0   | 0  | 0.5 | 0.047 | 0.029 |
| 250  | 0   | 0  | 0   | 0.047 | 0.053 |
| 250  | 75  | 35 | 0   | 0.052 | 0.060 |
| 250  | 0   | 0  | 0.5 | 0.063 | 0.054 |
| 250  | 75  | 35 | 0.5 | 0.061 | 0.030 |
| 500  | 0   | 0  | 0   | 0.048 | 0.053 |
| 500  | 75  | 35 | 0   | 0.056 | 0.047 |
| 500  | 150 | 70 | 0   | 0.052 | 0.048 |
| 500  | 0   | 0  | 0.5 | 0.053 | 0.051 |
| 500  | 75  | 35 | 0.5 | 0.054 | 0.035 |
| 500  | 150 | 70 | 0.5 | 0.050 | 0.047 |
| 1000 | 0   | 0  | 0   | 0.043 | 0.051 |
| 1000 | 75  | 35 | 0   | 0.043 | 0.046 |
| 1000 | 150 | 70 | 0   | 0.053 | 0.039 |
| 1000 | 0   | 0  | 0.5 | 0.050 | 0.038 |
| 1000 | 70  | 35 | 0.5 | 0.046 | 0.042 |
| 1000 | 150 | 70 | 0.5 | 0.062 | 0.041 |

Table 2. Power to detect heterogeneity in treatment effects, based on the linear regression model and the Random Forest predictions from the ALS example with an effect size of 0.19

| Model Prediction | Sample Size | Number of Nuisance Variables | Effect Distribution | | | | | |
|---|---|---|---|---|---|---|---|---|
| | | | Spread | 90/10 Cont. | 75/25 Cont. | 50/50 Cont. | 25/75 Cont. | 90/10 Bin. |
| Linear Regression | 3,000 | 0 | 1 | 1 | 1 | 1 | 1 | 1 |
| | | 20 | 1 | 1 | 1 | 1 | 1 | 0.958 |
| | | 50 | 0.924 | 1 | 0.998 | 0.968 | 0.988 | 0.878 |
| | | 100 | 0.256 | 0.002 | 0.002 | 0.002 | 0.498 | 0.658 |
| | 1,000 | 0 | 0.908 | 1 | 1 | 1 | 0.996 | 0.648 |
| | | 20 | 0.294 | 0.062 | 0.058 | 0.098 | 0.496 | 0.392 |
| | | 50 | 0.008 | 0 | 0 | 0 | 0.006 | 0.182 |
| | | 100 | 0 | 0 | 0 | 0 | 0 | 0.112 |
| Random Forest | 3,000 | 0 | 0.96 | 1 | 1 | 0.97 | 1 | 0.99 |
| | | 20 | 0.94 | 0.91 | 0.91 | 0.90 | 0.97 | 0.91 |
| | | 50 | 0.46 | 0.28 | 0.32 | 0.29 | 0.47 | 0.26 |
| | | 100 | 0.01 | 0.00 | 0.00 | 0.02 | 0.01 | 0.00 |
| | 1,000 | 0 | 0.23 | 0.08 | 0.06 | 0.10 | 0.44 | 0.42 |
| | | 20 | 0.16 | 0.03 | 0.06 | 0.09 | 0.19 | 0.12 |
| | | 50 | 0 | 0 | 0 | 0 | 0 | 0 |
| | | 100 | 0 | 0 | 0 | 0 | 0 | 0 |

Table 3. Power to detect heterogeneity in treatment effects, based on the linear regression model and the Random Forests predictions from the ALS example with an effect size of 0.38

| Model Prediction | Sample Size | Number of Nuisance Variables | Effect Distribution | | | | | |
|---|---|---|---|---|---|---|---|---|
| | | | Spread | 90/10 Cont. | 75/25 Cont. | 50/50 Cont. | 25/75 Cont. | 90/10 Bin. |
| Linear Regression | 3,000 | 0 | 1 | 1 | 1 | 1 | 1 | 1 |
| | | 20 | 1 | 1 | 1 | 1 | 1 | 1 |
| | | 50 | 1 | 1 | 1 | 1 | 1 | 1 |
| | | 100 | 1 | 1 | 1 | 1 | 1 | 1 |
| | 1,000 | 0 | 1 | 1 | 1 | 1 | 1 | 1 |
| | | 20 | 1 | 1 | 1 | 1 | 1 | 0.996 |
| | | 50 | 0.682 | 0.106 | 0.108 | 0.158 | 0.582 | 0.944 |
| | | 100 | 0 | 0 | 0 | 0 | 0 | 0.706 |
| Random Forest | 3,000 | 0 | 1 | 1 | 1 | 1 | 1 | 1 |
| | | 20 | 1 | 1 | 1 | 1 | 1 | 1 |
| | | 50 | 1 | 1 | 1 | 1 | 1 | 1 |
| | | 100 | 0.81 | 0.60 | 0.55 | 0.52 | 0.66 | 0.74 |
| | 1,000 | 0 | 1 | 1 | 1 | 1 | 1 | 0.99 |
| | | 20 | 0.97 | 0.94 | 0.98 | 0.98 | 0.99 | 0.99 |
| | | 50 | 0.29 | 0.18 | 0.11 | 0.16 | 0.24 | 0.24 |
| | | 100 | 0 | 0 | 0 | 0 | 0 | 0 |

APPENDIX

A. Parameter estimates from the linear regression model using the ALS example with 17 selected covariates.

| Covariates | Coefficients of the Treatment Group | Coefficients of the Control Group |
| --- | --- | --- |
| (Intercept) | -3.56945 | -2.88826 |
| Delta Flag | 0.01969 | 0.17810 |
| Respiratory Rate | -0.00099 | 0.01067 |
| Temperature | 0.10752 | 0.09247 |
| Weight(kg) | 0.00226 | 0.00166 |
| Height(cm) | -0.00555 | -0.00442 |
| Diastolic Blood Pressure | -0.00311 | -0.00204 |
| Systolic Blood Pressure | 0.00110 | -0.00113 |
| Pulse | -0.00365 | -0.00431 |
| Gender | 0.00712 | -0.03439 |
| Age | 0.00130 | -0.00327 |
| White | 0.03524 | -0.01493 |
| severity | -0.04591 | -0.06893 |
| Diagnosis Delta | -0.00036 | -0.00022 |
| Limb Only | -0.08336 | 0.08838 |
| Bulbar Only | -0.33667 | -0.08348 |
| Start Delta | -0.00019 | -0.00037 |
| Use Riluzole | -0.07150 | -0.22549 |

B. Means and SDs of the continuous covariates in the ALS example

|  | Mean | SD | Median | Min. | Max. |
| --- | --- | --- | --- | --- | --- |
| Systolic Blood Pressure | 131.88 | 16.63 | 130 | 85 | 206 |
| Age | 54.70 | 11.35 | 55 | 18 | 80 |
| Respiratory Rate | 17.19 | 3.27 | 16 | 6 | 42 |

C. Distribution of binary covariates in the ALS example

|  | Category | N | Percentage |
| --- | --- | --- | --- |
| Delta Flag (longer than 1 year) | Yes | 93 | 3.20% |
|  | No | 2817 | 96.80% |
| Limb Only | Yes | 1952 | 67.08% |
|  | No | 958 | 32.92% |
| Gender | Male | 1848 | 63.51% |
|  | Female | 1062 | 36.49% |
| Use Riluzole | Yes | 1112 | 38.21% |
|  | No | 1798 | 61.79% |